\documentstyle[12pt]{article}

\begin{document}
\begin{center}
{\Large{\bf Probability representation of quantum mechanics and star product quantization }}
\end{center}
\medskip
{\bf
S. N. Belolipetskiy$^2$, V. N. Chernega$^1$, O. V. Man'ko$^{1,2}$, V. I. Man'ko$^{1,3,4}$}\\
\medskip
{\it $^1$Lebedev Physical Institute, Russian Academy of Sciences\\
Leninskii Prospect 53, Moscow 119991, Russia\\
\smallskip
$^2$Bauman Moscow State Technical University\\
The 2nd Baumanskaya Street 5, Moscow 105005, Russia\\
$^3$ -  Moscow Institute of Physics and Technology (State University),  Institutskii per. 9, Dolgoprudnyi, Moscow Region 141700, Russia\\
$^4$ -  Tomsk State University, Department of Physics, Lenin Avenue 36, Tomsk 634050, Russia, Lebedev Physical Institute, Leninskii Prospect 53, Moscow 119991, Russia}\\
\smallskip
$^*$Corresponding author e-mail:~~~mankoov\,@\,lebedev.ru

\begin{abstract}\noindent
The review of star-product formalism providing the possibility to describe quantum states and quantum observables by means of the functions called symbols of operators which are obtained by means of bijective maps of the operators acting in Hilbert space onto these functions is presented. Examples of the Wigner-Weyl symbols (like Wigner quasi-distributions) and tomographic probability distributions (symplectic, optical and photon-number tomograms) identified with the states of the quantum systems are discussed. Properties of quantizer-dequantizer operators which are needed to construct the  bijective maps of two operators (quantum observables) onto the symbols of the operators are studied. The relation of the structure constants of the associative star-product of the operator symbols to the quantizer-dequantizer operators is reviewed. 
\end{abstract}
\medskip


\section{Introduction}
In classical mechanics and in classical statistical physics as well as in electromagnetic field theory the observables are described by the functions of positions and momenta like energy of a particle or of the spatial coordinates and time like strength of electric or magnetic field. The product of these functions is standard point-wise product which is commutative and associative. It means that the product of two functions, e. g. $f_1(q,p)$ and $f_2(q,p)$ gives the function $F_{12}(q,p)=f_1(q,p)\dot f_2(q,p)$ and $F_{21}(q,p)=F_{12}(q,p)$ as well as the product of three functions satisfies the associativity condition $\left(f_1(q,p)f_2(q,p)\right)f_3(q,p)=f_1(q,p)\left(f_2(q,p)f_3(q,p)\right)$. If one has the function $f({\bf x})$ the product can be expressed by means of the formula
\begin{equation}\label{eq.1o}
f_1({\bf x})f_2({\bf x})=\int\delta({\bf x}-{\bf x}_1)\delta({\bf x}-{\bf x}_2)f_1({\bf x}_1)f_2({\bf x}_2)\,d{\bf x}_1\,d{\bf x}_2,
\end{equation}
where $\delta({\bf y})$ with ${\bf y}={\bf x}-{\bf x}_k$, $k=1,2$ is Dirac delta function. The formula can be rewritten in generic form
\begin{equation}\label{eq.2o}
(f_1\ast f_2)({\bf x})=\int K({\bf x} _1,{\bf x}_2,{\bf x})f_1({\bf x}_1)f_2({\bf x}_2)\,d {\bf x}_1d{\bf x}_2.
\end{equation}
where $K({\bf x}_1,{\bf x}_2,{\bf x})$ is the kernel of the form 
\begin{equation}\label{eq.3o}
K({\bf x}_1,{\bf x}_2,{\bf x})=\delta({\bf x}-{\bf x}_1)\delta({\bf x}-{\bf x}_2).
\end{equation}
This kernel provides the properties of the point-wise products commutativity and associativity. The generic star-product of the functions which is not commutative but associative can be described by other kinds of kernels called structure constants of the associative product. If the variables ${\bf x}$ take discrete values either in finite or infinite domains the Dirac delta  functions in (\ref{eq.1o}), (\ref{eq.2o}), (\ref{eq.3o}) are replaced by Kronecker delta symbols and integral in (\ref{eq.2o}) is replaced by sum over the discrete variables. In quantum mechanics and quantum field theory the states and observables are associated with operators. For example the quantum oscillator position is described by position operator $\hat q$ which acts on the wave function $\psi(x)$ in position representation as $\hat q\psi(x)=x \psi(x)$. The momentum of the system is described by the operator $\hat p$ which acts on the wave function as $\hat p\psi(x)=-i\hbar\frac{\partial\psi(x)}{\partial x},$  where $\hbar$  is Planck constant. The product of position and momentum is non-commutative which is expressed as property of commutator for these physical observables $[\hat p,\hat q]=\hat p\hat q-\hat q\hat p=-i\hbar\hat 1.$ This property provides the uncertainty relation for position and momentum by Heisenberg \cite{Heisenberg1927} and by  Schr\"odinger \cite{Schrodinger1930}  and Robertson \cite{Robertson1930}. The physical meaning of the uncertainty relation is contained in the fact that one can not measure position and momentum of the oscillator simultaneously. The idea to introduce the formalism of star-product to describe the quantum phenomenon is: to find the bijective map of operators onto the functions. Then one can make the calculations using these functions. The product of operators is known to be associative, i.e. $(\hat A\hat B)\hat C=\hat A(\hat B\hat C)$. If one constructs the map of the functions associated with the operators the functions must be multiplied according to the formulas (\ref{eq.2o}) with some kernels which are not given in the form (\ref{eq.3o}) but have another form to violate the commutativity but preserve the associativity conditions. 

The aim of our work is to give the review of available different kinds of the star-product of the functions and to construct a generic scheme of quantizer-dequantizer operators used to find the kernels of star-products. In fact, the different forms of star-product of the functions correspond to different representations of the operators identified with the operators of physical observables. The different aspects of mathematical formalism of quantum mechanics were discussed in \cite{Mario}  and \cite{Vourdas}. In \cite{Mario} was shown that there exists a bijective relation between monotone quantum metrics associated with different operator monotone functions. In \cite{Vourdas} various Euclidean, hyperbolic and elliptic analytic representations are introduced and relations among them are discussed. 

The work is organized as follows. In Section 2 following \cite{Marmo,Krakov,Patricia} we review general scheme of quantization based on star-product of functions without concrete realization of the map of operators onto functions and discuss the expression for the kernel of star-product in terms of quantizaer-dequantizer operators. In Section 3 we discuss the quantum observables and their connection with dual quantization scheme. In Section 4 we review  the notion of  Wigner function and the quantizer-dequantizer operators for the Wigner-Weyl symbols of the operators. In Section 5 we discuss the symplectic and optical tomographic state representation. In Section 6 we review the tomographic symbols of operators, dual tomographic symbols of operators and mean values of quantum observables in the framework of star-product quantization scheme. In Section 7 we review the photon number tomography as an example of star-product quantization scheme. In Section 8 we discuss the relation of optical tomogram and photon-number tomogram. In Section 9 we review classical tomographic symbols. In Section 10 we discuss the star-product kernel of classical tomographic symbols and its connection with quantum ones. In Sec. 11 we give perspectives and conclusions.

\section{General scheme}
\noindent In quantum mechanics, observables are described by operators acting in the Hilbert space of states. In classical mechanics they are described by $c$-number functions. Let us rise a problem, how to describe quantum observable in classical--like way. 
In order to solve this problem we review first a general construction and provide general relations and properties of a map of operators onto functions without a concrete realization of the map. Following \cite{Marmo}, \cite{Krakov}, \cite{Patricia}, let us consider an operator $\hat A$ acting in a Hilbert space. We construct the function $f_{\hat A}({\bf x})$ of vector variables ${\bf x}=(x_1,x_2,\ldots,x_n)$, supposing that we have a set of operators $\hat U({\bf x})$ acting in Hilbert space such that 
\begin{equation}\label{eq.1}
f_{\hat A}({\bf x})=\mbox{Tr}\left[\hat A\hat{\cal U}({\bf x})\right].
%=\sum A_{m_1m_2}^{(j)}U_{m_2m_1}^{(j)}.
\end{equation}
The operator $\hat{\cal U}({\bf x})$ is called dequantizer \cite{Patricia} due to it maps operator $\hat A$ onto the function. The function $f_{\hat A}({\bf x})$ is called symbol of the operator $\hat A$. Let us suppose that operators $\hat{\cal D}({\bf x})$ acting in Hilbert  space exist  and the relation~(\ref{eq.1}) has an inverse. The map from function to operator is of the form
\begin{equation}\label{eq.2}
\hat A= \int f_{\hat A}({\bf x})\hat {\cal D}({\bf x})~d{\bf x}. 
\end{equation}
The operator $\hat{\cal D}({\bf x})$ is called quantizer \cite{Patricia}. We have the bijective map 
\begin{equation}\label{eq.3}
f_{\hat A}({\bf x})\leftrightarrow\hat A.
\end{equation}
Formulas (\ref{eq.1}) and (\ref{eq.2}) are self-consistent if the following property of the quantizer and dequantizer exists
\begin{equation}\label{eq.4}
\mbox{Tr}[\hat {\cal U}({\bf x})\hat{\cal D}({\bf x'})]=\delta({\bf x}-{\bf x'}).
\end{equation}
The delta-function in (\ref{eq.4}) is used in the case of continuous variables ${\bf x}$, and the Kronecker symbol instead of Dirac delta-function is used in the case of discrete variable ${\bf x}$.
The introduced map provides the nonlocal associative product  (star-product) of two functions (symbols of operators). The product of two symbols  $f_{\hat A}({\bf x})$ and $f_{\hat B}({\bf x})$
corresponding to two operators $\hat A$ and
$\hat B$ with the map (\ref{eq.1}), (\ref{eq.2}) can be introduced in the form (see, e.g. \cite{Marmo}) 
\begin{equation}\label{eq.5}
f_{\hat A\hat B}({\bf x})=f_{\hat A}({\bf x})\ast
f_{\hat B} ({\bf x})=\int f_{\hat A}({\bf x'})f_{\hat B} ({\bf x''})K({\bf x'},{\bf x''},{\bf x})d{\bf x'}d{\bf x''},
\end{equation}
where the kernel of star product $K({\bf x'},{\bf x''},{\bf x})$ is of the form
\begin{equation}\label{eq.6}
K({\bf x'},{\bf x''},{\bf x})=\mbox{Tr}[\hat {\cal D}({\bf x''})\hat{\cal  D}({\bf x'})\hat{\cal U}({\bf x})].
\end{equation}
One can see that the kernel of the star-product (\ref{eq.6}) is linear with respect to the dequantizer and nonlinear in the quantizer operator. 
Since the standard product of operators in a Hilbert space is an associative product 
$\hat A(\hat B \hat C)=(\hat A\hat B)\hat C$, 
the product of functions (symbols of operators) has to be associative too 
\begin{equation}\label{eq.6a}
f_{\hat A}({\bf x})\ast\Big(f_{\hat B}({\bf x})
\ast f_{\hat C}({\bf x})\Big)=
\Big(f_{\hat A}({\bf x})\ast f_{\hat B}({\bf x})\Big)
\ast f_{\hat C}({\bf x}).
\end{equation}
The associativity condition for functions (symbols of operators) means that the kernel of star-product (\ref{eq.6}) satisfies the nonlinear equation (see e.g.\cite{Patricia})
\begin{eqnarray}\label{patC2}
\int K({\bf x}_1,{\bf x}_2, {\bf y})K({\bf y},{\bf x}_3, {\bf
x}_4)d{\bf y}=
\int K({\bf x}_1,{\bf y}, {\bf x}_4)K({\bf x}_2,{\bf x}_3, {\bf
y})d{\bf y}.
%\nonumber
\end{eqnarray}
If we take as operator $\hat A$ the density operator $\hat\rho$ determining some quantum state then the function $f_\rho({\bf x})$ (symbol of operator $\hat\rho$) also determines this quantum state. In \cite{Marmo4} the quantizer-–dequantizer formalism was used to describe the evolution of a quantum system and it was shown that if the set of states is invariant with respect to some unitary evolution, the quantizer–-dequantizer provides a classical-like realization of the system dynamics.

\section{The dual star--product scheme and quantum
observable}
Let us consider the following map 
\begin{equation}\label{SPeq.1dual}
f^{(d)}_{\hat A}({\bf x})=\mbox{Tr}\left[\hat A\hat{\cal D}({\bf
x})\right],
\end{equation}
\begin{equation}\label{SPeq.2dual} \hat A= \int f^{(d)}_{\hat
A}({\bf x})\hat{\cal U}({\bf x})~d{\bf x}.
\end{equation}
We permute the quantizer and the dequantizer. It is possible because the compatibility condition (\ref{eq.6}) is valid in
both cases. We consider the new pair of quantizer--dequantizer as dual pair to the initial one
\begin{eqnarray*}
\hat{\cal U}^\prime({\bf x})\Longrightarrow \hat{\cal D}({\bf
x}),\quad \hat {\cal D}^\prime({\bf x})\Longrightarrow\hat{\cal
U}({\bf x}).
\end{eqnarray*}
This interchange is possible due to a specific symmetry of the equation 
(\ref{patC2}) for associative star--product kernel. The
star--product of dual symbols~$f^{(d)}_{\hat A}({\bf x})$,
$f^{(d)}_{\hat B}({\bf x})$ of two operators $\hat A$ and $\hat B$
is described by dual integral kernel
\begin{eqnarray}\label{kerneldual}
K^{(d)}({\bf x}'',{\bf x}',{\bf x})=\mbox{Tr}\left[\hat{\cal
U}({\bf x}'')\hat{\cal U}({\bf x}')\hat{\cal D}({\bf x})\right].
\end{eqnarray}
The dual kernel (\ref{kerneldual}) is another solution of
nonlinear equation (\ref{patC2}) \cite{Patricia}. Let us consider the mean value of quantum observable which is determined by the operator $\hat A$
\begin{eqnarray*}
&&\langle \hat A\rangle=\mbox{Tr}\big(\hat\rho\hat A\big)= \int
f_{\hat \rho}({\bf x}) \mbox{Tr}\big(\hat{\cal D}(\bf x)\hat A
\big) d~\bf x\nonumber.
\end{eqnarray*}
Using the expression for dual symbol of operator(\ref{SPeq.1dual}) we obtain the formula
\begin{eqnarray*}
%&&
\langle\hat A\rangle=%\\
%&&\\
%&&
\int f_\rho(\bf x)f_{\hat A}^{(d)}(\bf x)\, d~\bf x.
\end{eqnarray*}
So, one can see, that the mean value of an observable $\hat A$ can be calculated as overlap integral of the tomographic symbol of 
the density operator $f_{\rho}(\bf x)$ in a given quantization scheme and the symbol $f_{\hat A}^{(d)}(\bf x)$ of the observable 
$\hat A$ in the dual scheme.

\section{Wigner-Weyl symbol}
In this section we will consider an example of Heisenberg-Weyl representation. Let us introduce the operators (we assume Planck constant $\hbar=1$)
\begin{equation}\label{eq.1w}
\hat {\cal U}(q,p)=\int_{-\infty}^{\infty}|q+ \frac{u}{2}\rangle\langle q-\frac{u}{2}| e^{-i p u}d u,
\end{equation}
and
\begin{equation}\label{eq.2w}
\hat{\cal D}(q',p')=\frac{1}{2\pi}\hat {\cal U}(q',p'),
\end{equation}
where
$|q+u/2\rangle$ is the eigenvector of position operators $\hat q$, i.e. $\hat q| x\rangle=x|x\rangle$ and it satisfies the condition
\begin{equation}\label{eq.4w}
\langle x|x'\rangle=\delta(x-x').
\end{equation}
Operator (\ref{eq.1w}) is dequantizer operator according to general scheme (\ref{eq.1}). Operator (\ref{eq.2w}) is quantizer operator according to (\ref{eq.2}) in the Wigner-Weyl star-product quantization scheme. 
One can check that
\begin{equation}\label{eq.5w}
\mbox{Tr}\left(\hat{\cal U}(q,p)\hat{\cal D}(q',p')\right)=\delta(q-q')\delta(p-p').
\end{equation}
In view of this property according to general scheme of constructions the symbols of the operators the function
\begin{equation}\label{eq.6w}
W_A(q,p)=\mbox{Tr}\left(\hat{\cal U}(q,p)\hat A\right)
\end{equation}
which is Wigner-Weyl symbol of the operator $\hat A$ determines the operator $\hat A$, i.e,
\begin{equation}\label{eq.7w}
\hat A=\frac{1}{2\pi}\int W_A(q,p)\hat U(q,p)\, d q d p.
\end{equation}
The kernel of star-product of the Wigner-Weyl symbols \cite{Groen} reads
\begin{equation}\label{eq.8w}
\mbox{Tr}\left(\hat{\cal D} (q_1,p_1)\hat{\cal D}(q_2,p_2)\hat{\cal U}(q_3,p_3)\right)=\frac{1}{2\pi^2}\exp\left[2i\left(q_1p_2-q_2p_1+q_2p_3-q_3p_2+q_3p_1-q_1p_3\right)\right].
\end{equation}
For example, the symbol of the oscillator ground state is 
\begin{equation}\label{eq.9w}
W_\rho(q,p)=W_\rho(q,p)=2\exp\left(-q^2-p^2\right).
\end{equation}
As it can be shown the operator $\hat{\cal U}$ can be given in the other form 
\begin{equation}\label{eq.8w}
\hat {\cal U}(q,p)=2\hat D(2\alpha)\hat I,
\end{equation}
where $\hat I$ is parity operator, $\alpha=\frac{q+i p}{\sqrt2}$, $\hat D(2\alpha)$ is displacement operator. The displacement operator
is expressed through creation and annihilation operators in the form 
\begin{equation}\label{eq.9w} 
\hat D({\alpha})=\exp(\alpha\hat a^{+}
-\alpha^{\ast}\hat a).
\end{equation}
where creation and annihilation operators are expressed through operators of position $\hat q$ and momentum $\hat p$ of oscillator 
\begin{equation}\label{eq.10w}
\hat a=\frac{\hat q+i\hat p}{\sqrt 2},\quad
\hat a^{+}=\frac{\hat q- i\hat p}{\sqrt2},
\end{equation}
The Wigner-Weyl star-product scheme is self-dual due to condition (\ref{eq.2w}). 
It is worthy to add that Weyl operators and symbols were investigated e.g. in \cite{Gousson}. A new quantum mechanical formalism based on the probability representation of quantum states is used to investigate the special case of the measurement problem, known as Schr\"odinger’s cat paradox and the EPR-paradox in \cite{Foukzon}. The evolution of the discrete Wigner function for prime and the power of prime dimensions using the discrete version of the star-product operation was investigated in \cite{Klimov1}. The explicit differential Moyal-like form of the star product is found and analyzed in the semi-classical limit in \cite{Klimov3}. In \cite{Anielo} the associative star product of functions was investigated with the help of a square integrable representation of a locally compact group.
The Wigner functions and tomographic probability distributions of two-qubit states were discussed in \cite{Adam}, where the kernel of the map, which provides the expression of the state tomogram in terms of the discrete Wigner function of the two-qubit state and the kernel of the inverse map and the connection of the constructed maps with the star-product quantization scheme is obtained  in an explicit form. In \cite{Tombesi}  the evolution of Werner-like mixture by considering two correlated but different degrees of freedom was introduced and its tomographic characterization was provided. The Wigner functions for the harmonic oscillator including corrections from generalized uncertainty principles and the corresponding marginal probability densities are studied in \cite{Das}. The investigation of general quantum mechanical commutation relations consistent with the Heisenberg evolution equations was studied in \cite{Kapustik}.  The review of a family of non-commutative star-products based on a Weyl map is done in \cite{Lizzi}.

\section{Notion of quantum state in symplectic and optical tomography approaches}
The symplectic tomography was introduced in~\cite{Mancini96}. Tomographic approach 
to quantum states that leads to a probability representation of quantum states was discussed in \cite{ManciniFP}, \cite{TombesiSemOpt}, \cite{my1}, \cite{Mancini2}. The state in symplectic tomography scheme is determined by probability distribution function $w(X,\mu,\nu)$, which is called symplectic tomogram. The generic linear 
combination of quadratures which is a measurable observable $\left(\hbar =1\right)$ is of the form 
\begin{equation}\label{X}
\widehat X=\mu \hat q+\nu\hat p\,,
\end{equation}
where $\hat q$ and $\hat p$ are the position and momentum, respectively, and real parameters $\mu$ and $\nu$ determine the reference frame in classical phase space. The symplectic tomogram $w\,(X,\,\mu,\,\nu )$
is nonnegative function which is normalized with respect to the variable $X$ (position). The physical meaning of the parameters $\mu $ and $\nu $ is that they describe an ensemble of rotated and scaled reference frames in which the position $X$ is measured. For $\mu =\cos \,\theta $ and $\nu =\sin \,\theta ,$ the symplectic tomogram coincides with  distribution for the homodyne-output variable used in optical tomography~\cite{BerBer}, \cite{VogRis} and named optical tomogram
\begin{equation}\label{eq.Opt}
w_{opt}(X,\theta)=w(X,\cos\theta,\sin\theta). \label{ot}
\end{equation}
The information contained in the symplectic tomogram $w\left(X,\,\mu,\,\nu \right)$ is overcomplete. To determine the quantum state completely, it is sufficient to give the function for arguments with the constraints $\left(\mu^2+\nu^2=1\right)$ which corresponds to the optical tomography scheme which is realized experimentally in \cite{RaymerPRL93}, \cite{Mlynek}, i.e., $\mu=\cos\theta$ and the rotation angle $\theta$ labels the reference frame 
in classical phase space. Symplectic tomogram can be reconstructed from optical tomogram using the relation 
\begin{equation}\label{eq.theta}
w(X,\mu,\nu)=\frac{1}{\sqrt{\mu^2+\nu^2}}w_{opt}\left(\frac{X}{\sqrt{\mu^2+\nu^2}},\mbox{arctg}\frac{\nu}{\mu}\right).
\end{equation}
The process of Stimulated Raman Scattering was investigated in the frame of the symplectic tomography representation in \cite{VRMB}. The quantum entanglement in Raman Scattering was investigated in \cite{Pathak}, \cite{Perina}.

\section{Symplectic tomography in the framework of star-product quantization}
The tomographic symbol $w_{\hat A}({\bf x})$ of the operator $\hat A$ is obtained by means of the dequantizer operator of the form 
\begin{eqnarray*}
\hat{\cal U}(X,\mu,\nu)=\delta(X\hat 1-\mu\hat q-\nu\hat p)
\end{eqnarray*}
where vector ${\bf x}=(X,\mu,\nu)$ has the arguments which are real numbers, $\hat 1$ is identity operator. The quantizer operator in symplectic tomography is 
\begin{eqnarray*}
\hat{\cal D}(X,\mu,\nu)=\frac{1}{2\pi} \exp\left(iX\hat1-i\nu\hat
p-i\mu\hat q\right).
\end{eqnarray*}
The kernel of star--product given by (\ref{eq.6}) of two tomographic symbols of operators $\hat A$ and $\hat B$  has the following form \cite{Marmo}, \cite{Krakov}
\begin{eqnarray*}
&&K(X_1,\mu_1,\nu_1,X_2,\mu_2,\nu_2,X\mu,\nu)=
\frac{\delta\Big(\mu(\nu_1+\nu_2)-\nu(\mu_1+\mu_2)\Big)}{4\pi^2}\nonumber\\
&&\times\exp\Big(\frac{i}{2}\Big\{\left(\nu_1\mu_2-\nu_2\mu_1\right)
+2X_1+2X_2-\frac{2(\nu_1+\nu_2)X}{\nu}\Big\}\Big).
\end{eqnarray*}
Let us consider the mean value of quantum observable $\hat A$
\begin{eqnarray*}
&&\langle \hat A\rangle=\mbox{Tr}\big(\hat\rho\hat A\big)=
\mbox{Tr}\int w\big(X,\mu,\nu\big)\hat{\cal D}(X,\mu,\nu)\hat A\,
d X\,d\mu\, d\nu=\nonumber\\
&&\int w\big(X,\mu,\nu\big)\mbox{Tr}\big(\hat{\cal D}(X,\mu,\nu)\hat
A \big), d X\,d\mu\, d\nu.
\end{eqnarray*}
For this purpose we introduce the dual tomographic symbol (\ref{SPeq.1dual}) in symplectic tomography scheme 
\begin{eqnarray*}
w^{(d)}_{\hat A}({X,\mu,\nu})=\mbox{Tr}\left[\hat A\hat{\cal D}(
X,\mu,\nu)\right].
\end{eqnarray*}
Then for the mean value of observable $\langle\hat A\rangle$ one has 
\begin{eqnarray*}
&&\langle\hat A\rangle=\int w\big(X,\mu,\nu\big)w_{\hat A}^{(d)}(X,\mu,\nu)\, d X\,d\mu\,
d\nu.
\end{eqnarray*}
The mean value of an observable $\hat A$ in symplectic tomography scheme is given by the overlap integral of the symplectic tomogram (the tomographic symbol of the density operator) $w\big(X,\mu,\nu\big)$ in the given  
quantization scheme and the symbol $w_{\hat A}^{(d)}(X,\mu,\nu)$ of the observable $\hat A$ in the dual scheme. 
The tomograms and the eigenvalues of energy are shown to be computed in terms of tomographic symbols in \cite{Bazrafkan}. The dynamic of quantum particles was described by the Kolmogorov equations for non-negative propagators in the tomography representation in \cite{Filinov}. The symmetrized product of quantum observables is defined in \cite{Slobodan}.

\section{Photon--number tomography as example of star--product quantization}
The photon-number tomography was introduced in \cite{vogel}, \cite{wodk}, \cite{euroPLMancini} and developed in \cite{JRLR2003}, \cite{myMexico}, \cite{Petrograd2004}, \cite{SPIE2005}. It is the method to reconstruct density operator of quantum state using measurable probability
distribution function (photon statistics) called photon-number tomogram. In photon--number tomography the discrete random variable is measured for reconstructing quantum state. The photon--number
tomogram
\begin{equation} \label{PNT}
\omega(n,\alpha)=\langle n\mid\hat
D(\alpha)\hat \rho\hat D^{-1}(\alpha)\mid n\rangle
\end{equation}
is the function of integer photon number $n$ and complex number $\alpha$, $\hat\rho$ is the state density operator. The photon--number tomogram is the photon distribution function (the 
probability to have $n$ photons) in the state described by the displaced density operator.  For example, %photon--number tomogram for oscillator ground state is
%\[
%w_0(n,\alpha)=\frac{e^{-|\alpha|^2}}{n!}|\alpha|^{2n}.
%\]
the photon--number tomograms of excited oscillator states with density operators $\hat\rho_m=\mid m\rangle\langle m \mid$ are
\begin{eqnarray}\label{PNT1}
&&w^{(m)}(n,\alpha)=\frac{n!}{m!}\mid\alpha\mid^{2(m-n)}e^{-\mid\alpha\mid^2}
\left(L_n^{m-n}(\mid\alpha\mid^2)\right)^2,\quad m\geq n\,,\nonumber\\
&&w^{(m)}(n,\alpha)=\frac{m!}{n!}\mid\alpha\mid^{2(n-m)}e^{-\mid\alpha\mid^2}
\left(L_m^{n-m}(\mid\alpha\mid^2)\right)^2,\quad m\leq n,\label{PNT1}
\end{eqnarray}
where $L_n^m(x)$ are Laguerre polynomials. In \cite{Kiukas} the state reconstruction from the measurement statistics of phase space observables generated by photon number states was considered.

Let us consider photon--number tomogram in the framework of star--product quantization following \cite{laserphys2009}, \cite{PhysicsScrFinland}, \cite{AIP2011}, \cite{AIP2012}. In the given photon--number tomography quantization scheme the dequantizer operator  is of the form
\begin{equation} \label{dequantizerphnumtom}
\widehat{\cal U}(\mbox{\bf x})=\hat D(\alpha)|n\rangle\langle
n|\hat D^{-1}(\alpha), \,\mbox{\bf x}=(n,\alpha).
\end{equation}
The quantizer operator in photon number tomography scheme is
\begin{equation}\label{quantizerphnumtom}
\widehat{\cal D}(\mbox{\bf x})=\frac{4}{\pi(1-s^2)}
\left(\frac{s-1}{s+1}\right)^{(\hat a^\dagger+\alpha^*)(\hat
a+\alpha)-n},
\end{equation}
where $s$ is ordering parameter \cite{Cah}, $\alpha$ is complex number 
$(\alpha=\mbox{Re}\,\alpha+i\,\mbox{Im}\,\alpha),$ 
$D(\alpha)$ is the Weyl displacement operator (\ref{eq.9w}). 
%\[\hat D(\alpha)=\exp(\alpha\hat a^{\dagger}-\alpha^*\hat a).\]
The kernel (\ref{eq.6}) of star--product of photon--number
tomograms in the photon--number tomography quantization
scheme is
\begin{equation}\label{kernelphotonnumbertommainscheme}
K(n_1,\alpha_1,n_2,\alpha_2,n_3,\alpha_3)=\mbox{Tr}\big[\hat{\cal
D}(n_1,\alpha_1)\hat{\cal D}(n_2,\alpha_2)\hat{\cal
U}(n_3,\alpha_3)\big].
\end{equation}
In explicit form it is 
\begin{eqnarray}
&&
K(n_1,\alpha_1,n_2,\alpha_2,n_3,\alpha_3)=\Big(\frac{4}{\pi(1-s^2)}
\Big)^2\exp\large(it(n_1+n_2-2n_3)\large)
\exp\large[-|-\alpha_3+\alpha_1\nonumber\\
&&-\alpha_1e^{-it} +\alpha_2
e^{-it}-\alpha_2e^{-2it}+\alpha_3e^{-2it}|^2 +\frac{1}{2}(\large(
-\alpha_3\alpha_1^*+\alpha_3^*\alpha_1-\alpha_1\alpha_2^*
+\alpha_1^*\alpha_2 -\alpha_2\alpha_3^*\nonumber\\
&&+\alpha_2^*\alpha_3
+\alpha_3\alpha_1^*e^{it^*}
 -|\alpha_1|^2e^{it^*}-\alpha_3\alpha_2^*e^{it^*}
+\alpha_1\alpha_2^*e^{it^*} -\alpha_3^*\alpha_1e^{-it}
 +|\alpha_1|^2e^{-it}\nonumber\\
&& +\alpha_3^*\alpha_2e^{-it}
-\alpha_1^*\alpha_2e^{-it}+\alpha_3\alpha_2^*e^{2it^*}
 -\alpha_1\alpha_2^*e^{2it^*}
+\alpha_1\alpha_2^*e^{-it+2it^*}
-|\alpha_2|^2e^{-it+2it^*}\nonumber\\
&&-|\alpha_3|^2e^{2it^*}
+\alpha_1\alpha_3^*e^{2it^*}
-\alpha_1\alpha_3^*e^{-it+2it^*}
 +\alpha_2\alpha_3^*e^{-it+2it^*}
-\alpha_3^*\alpha_2e^{-2it}
 +\alpha_1^*\alpha_2e^{-2it}\nonumber\\
&&-\alpha_1^*\alpha_2e^{it^*-2it}
 -|\alpha_2|^2e^{it^*-2it}
+|\alpha_3|^2e^{-2it}-\alpha_1^*\alpha_3e^{-2it}
 +\alpha_1^*\alpha_3e^{it^*-2it}
-\alpha_2^*\alpha_3e^{it^*-2it}\large)\large]\nonumber\\
&&\times L_{n_3}\large(|-\alpha_3+\alpha_1-\alpha_1e^{-it}+\alpha_2
e^{-it}-\alpha_2e^{-2it}+\alpha_3e^{-2it}|^2\large),\label{kernephnum}
\end{eqnarray}
where
\[\frac{s-1}{s+1}=e^{it}\]
and $L_n(x)$ is Laguerre polynomial. The kernel (\ref{kernephnum}) is solution of equation (\ref{patC2}).

\section{The relation of optical tomogram and photon number tomogram}
In this section we find the relations among optical, symplectic and
photon numbers tomograms of quantum state following \cite{laserphys2009}. The photon number
tomogram can be expressed in terms of the symplectic tomogram by
using the integral transform \cite{PhysicsScrFinland}
\begin{equation}\label{p1}
\omega(n,\alpha)= \int w(X,\mu,\nu) K(X,\mu,\nu,n,\alpha)\, dX\,
d\mu\,d\nu.
\end{equation}
Here the kernel of the integral transform is expressed in terms of
matrix elements of the displacement operator
\begin{equation}
K(X,\mu,\nu,n,\alpha)=\frac{1}{2\pi}\langle n\mid\hat
D(-\alpha)e^{i(X-\mu\hat q-\nu\hat p)}\hat D(\alpha)\mid n\rangle.
\end{equation}
The explicit dependence of the kernel on the real parameters of the
symplectic transform is given by the expression
\begin{equation}\label{p2}
K(X,\mu,\nu,n,\alpha)=\frac{1}{2\pi}\exp\left[iX+\frac{\nu-i\mu}{\sqrt2}\alpha^\ast
-\frac{\nu+i\mu}{\sqrt2}\alpha\right]\langle n\mid\hat
D\left(\frac{\nu-i\mu}{\sqrt 2}\right)n\rangle.
\end{equation}
Using the known formula for the diagonal elements of the
displacement operator
\begin{equation}\label{p3}
D_{n n}(\gamma)=\langle n\mid\hat D(\gamma)\mid
n\rangle=e^{-\frac{\mid\gamma\mid^2}{2}}L_n(\mid\gamma\mid^2),
\end{equation}
where $\gamma=(\nu+i\mu)/\sqrt 2$, one has the kernel expressed in terms of Laguerre polynomial
\begin{equation}\label{p4}
K(X,\mu,\nu,n,\alpha)=\frac{1}{2\pi}\exp\left[iX+\frac{\nu-i\mu}{\sqrt2}\alpha^\ast
-\frac{\nu+i\mu}{\sqrt2}\alpha\right]L_n\left(\frac{\nu^2+\mu^2}{2}\right).
\end{equation}
In view of this the photon number tomogram is expressed in terms of optical tomogram as follows
\begin{eqnarray}\label{p5}
w(n,\alpha)=&&\frac{1}{2\pi}\int_{0}^{2\pi}
\int_0^{\infty}\int_{-\infty}^{\infty}k\exp\left[i k\left(X-\sqrt
2(\alpha_1\cos\theta+\alpha_2\sin\theta)\right)
-\frac{k^2}{4}\right]\nonumber\\
&&\times L_n\left(\frac{k^2}{2}\right)w_0(X,\theta)\, d\theta\, d
k\, d X.
\end{eqnarray}
Here $\alpha$ is the complex number($\alpha=\alpha_1+i\alpha_2$). 
Let us introduce the characteristic function
\begin{equation}\label{P6}
F(k,\theta)=\int e^{i k X}w_0(X,\theta)\, d X=\langle e^{i k
X}\rangle, \quad k\geq 0.
\end{equation}
Then the photon number tomogram reads
\begin{eqnarray}
w(n,\alpha)=&&\frac{1}{2\pi}\sum_m\int_{0}^{2\pi}
\int_0^{\infty}k\exp\left[-i k\sqrt
2\left(\alpha_1\cos\theta+\alpha_2\sin\theta\right)
-\frac{k^2}{4}\right]\nonumber\\
&&\times L_n\left(\frac{k^2}{2}\right)(i)^m\frac{k^m}{m!}\langle
X^m\rangle_\theta\, d k\, d\theta.\label{p7}
\end{eqnarray}
Here we introduce moments of optical tomogram $\langle X^m\rangle_\theta=\int w(X,\theta)X^m \,d X.$  Let us consider an example of excited oscillator state 
$\hat\rho_m=\mid m\rangle\langle m \mid.$ One can show that the tomogram of this state reads
\begin{equation}\label{p13}
w_m(X,\mu,\nu)=\frac{e^{-\frac{X^2}{\sqrt{\mu^2+\nu^2}}}}
{\sqrt{\pi(\mu^2+\nu^2)}}\frac{1}{m!}\frac{1}{2^m}
H^2_m\left(\frac{X}{\sqrt{\mu^2+\nu^2}}\right).
\end{equation}
Applying general relation (\ref{p1}) we get for $m\leq n$
\begin{eqnarray}\label{p14}
&&\frac{m!}{n!}\mid\gamma\mid^{2(n-m)}e^{-\mid\gamma\mid^2}
\left(L_m^{n-m}(\mid\gamma\mid^2)\right)^2=\int
\frac{1}{2\pi}\exp\left[i X+\frac{\nu-i\mu}{\sqrt2}\gamma^\ast
-\frac{\nu+i\mu}{\sqrt2}\gamma\right]\nonumber\\&& \times
e^{-\frac{\nu^2+\mu^2}{4}}L_n(\frac{\nu^2+\mu^2}{2})
\frac{e^{-\frac{X^2}{\sqrt{\mu^2+\nu^2}}}}
{\sqrt{\pi(\mu^2+\nu^2)}}\frac{1}{m!}\frac{1}{2^m}
H^2_m\left(\frac{X}{\sqrt{\mu^2+\nu^2}}\right)d X\, d\mu\, d\nu
\end{eqnarray}
which provides a new integral relation of Hermite and Laguerre
polynomials \cite{laserphys2009}, \cite{PhysicsScrFinland}. For ground state with optical tomogram
\[ w_0(X,\theta)=\frac{-e^{X^2}}{\sqrt{\pi}}\]
one has the photon number tomogram
\[
w_0(n,\alpha)=\frac{e^{-|\alpha|^2}}{n!}|\alpha|^{2n}.
\]
which is Poissonian distribution. It is necessary to add that optical tomograms of the time-evolved states generated by the evolution of different kinds of initial wave packets in a Kerr medium and optical tomograms of maximally entangled states generated at the output modes of a beam splitter was theoretically studied in \cite{Rohith}, \cite{Rohith1}. In \cite{Facchi} a review of the Radon transform and the instability of the tomographic reconstruction process were discussed.

\section{ Classical Tomographic Symbols}
In this section we review following \cite{my1}, \cite{2004JRLR}, \cite{JRLR2001}, \cite{hlimit}, \cite{Pilyavets} the consideration of classical mechanics within the framework of the tomographic representation. The reversible relationship between the tomogram 
$w_f(X,\mu,\nu)$ of the probability distribution $f(q,p)$ in
classical mechanics was defined in~\cite{Pilyavets} as follows:
\begin{eqnarray}%\label{wf}
&&w_f(X,\mu,\nu)=\frac1{2\pi}\int f(q,p)\delta(X-\mu q-\nu p)
\,dq\,dp,\nonumber\\
%&&\\
&&f(q,p)=\frac1{2\pi}\int w_f(x,\mu,\nu)e^{i(X-\mu q-\nu
p)}\,dX\,d\mu\, d\nu.\label{wf}
%\nonumber
\end{eqnarray}
Here we assumed the normalization condition for $f(q,p)$ to be
$$\frac{1}{2\pi}\int f(q,p)\,dq\,dp = 1 $$
analogously to the normalization condition of the Wigner function
$W(q,p)$ in quantum mechanics~\cite{Wig32}.

According to~\cite{2004JRLR} in classical mechanics one can
introduce the operators $\hat A_{\rm cl}$ for which their formal
Weyl symbol $W_{A_{\rm cl}}(q,p)$ coincides with a classical
observable $A(q,p)$, i.e.,
\[
W_{A_{\rm cl}}(q,p)=2\,\mbox{Tr}\,\hat A_{\rm cl}\hat D(2\alpha)\hat
I=A(q,p).
\]
Thus we interpret a function in the phase space as Wigner--Weyl symbol of an operator $\hat A_{cl}$ acting in Hilbert space. Consequently, we can consider the phase-space function $A(q,p)$ as the classical Weyl symbol of the
corresponding observable in classical mechanics. The quantum tomographic symbol for a unity operator was found in~\cite{Shchukin1}. Due to mentioned above, we have the same result for the classical tomographic symbol, i.e.,
\begin{equation}
w_1(X,\mu,\nu)=-\pi|X|\delta(\mu)\delta(\nu). \label{w1}
\end{equation}
Since in quantum mechanics Weyl symbols for the position operator
$\hat q$ and momentum operator $\hat p$ are $c$-numbers $q$ and $p$,
in view of Eqs.~(\ref{wf}), we have for both classical and quantum
tomographic symbols:
\begin{eqnarray*}
&&w_q(X,\mu,\nu)=%\frac1{(2\pi)^2}\int qe^{ik(X-\mu q-\nu
\frac{\pi}2X|X|\delta'(\mu)\delta(\nu),\nonumber\\
&& w_p(X,\mu,\nu)=
\frac{\pi}2X|X|\delta(\mu)\delta'(\nu).\label{symbols}
\end{eqnarray*}

\section{Star-product kernel of classical tomographic symbols}
In this Section we present the kernel of star-product of functions-symbols of operators in classical mechanics and show its connection with star-product of quantum tomographic symbols of operators. The star-product kernel 
$K(X,\mu,\nu,X_1,\mu_1,\nu_1,X_2,\mu_2,\nu_2)$ for two tomographic
symbols $w_1(X_1,\mu_1,\nu_1)$ and $w_2(X_2,\mu_2,\nu_2)$ of observables was
defined in~\cite{2004JRLR} as follows:
\begin{eqnarray}
w(X,\mu,\nu)&=&\int
K(X,\mu,\nu,X_1,\mu_1,\nu_1,X_2,\mu_2,\nu_2)w_1(X_1,\mu_1,\nu_1)\nonumber\\
&&\times w_2(X_2,\mu_2,\nu_2)\,d
X_1\,d\mu_1\,d\nu_1\,dX_2\,d\mu_2\,d\nu_2, \label{integral}
\end{eqnarray}
where by the definition
$$w(X,\mu,\nu)=w_1(X,\mu,\nu)\ast  w_2(X,\mu,\nu).$$
To find the explicit form of tomographic star-product kernel
$K(X,\mu,\nu,X_1,\mu_1,\nu_1,X_2,\mu_2,\nu_2)$, let us substitute in
the definition of tomographic symbol the point-wise product of
functions $A(q,p)$ and $B(q,p)$ and then express these functions
(classical Weyl symbols) through its tomographic symbols. This leads
to the result for the commutative star-product kernel of the
tomographic symbols:
\begin{equation}
K(X,\mu,\nu,X_1,\mu_1,\nu_1,X_2,\mu_2,\nu_2) =
\frac{1}{(2\pi)^2}e^{i\left(X_1+X_2-X({\nu_1+\nu_2})/{\nu}\right)}
\delta\Big(\nu(\mu_1+\mu_2)-\mu(\nu_1+\nu_2)\Big).
\label{ClassicKernel}
\end{equation}
Analogously we can find the star-product kernel for quantum
tomographic symbols~\cite{Marmo} if we consider $W_A(q,p)$ and
$W_B(q,p)$ as usual quantum Weyl symbols of the operators $\hat A$
and $\hat B$ with the Weyl noncommutative star-product. The
relationship between tomographic star-product kernels in quantum and
classical mechanics reads
\[
K_{\mbox{\scriptsize{quant}}}(X,\mu,\nu,X_1,\mu_1,\nu_1,X_2,\mu_2,\nu_2)
=
K_{\mbox{\scriptsize{classic}}}(X,\mu,\nu,X_1,\mu_1,\nu_1,X_2,\mu_2,\nu_2)e^{
[i\left(\mu_2\nu_1-\mu_1\nu_2\right)/2]}.
\]
The relation between the quantum state description and the classical state description is elucidated in \cite{Ibort}. In \cite{Khrennikov} Brownian motion was considered in the space of fields and was shown that the corresponding probability distribution can be approximately described by the same mathematical formalism as is used in quantum mechanics and in theory of Hermitian operators in complex Hilbert space. In \cite{KhrennikobAlo} the review of classical probability representations of quantum states and observables is done and is shown that the correlations of the observables involved in the Bohm--Bell type experiments can be expressed as correlations of classical random variables.
The quantum--classical limits for quantum tomograms are studied and compared with the corresponding classical tomograms in \cite{Marmo3}. In \cite{Stornaiolo} a tomographic representation of quantum cosmology, in which tomograms (standard positive probability distribution functions) describe the quantum state of universe instead of the wave functions or density matrices was considered. The extension of the tomographic maps to the quantum case and a Weyl-Wigner quantization in  the classical case are considered in \cite{Facci2}. In \cite{VALLONE}  superspace (instead of Hilbert space) was employed in order to describe time evolution of density matrices in terms of path integrals, which are formally identical for quantum and classical mechanics. In \cite{elze} an alternative theory of hybrid classical-quantum dynamics based on notions of phase space was proposed. In \cite{Bondar} the classical limit  of the Wigner function was considered and was discussed its  transform into a classical Koopman--von Neumann wave function rather than into a classical probability distribution.

\section{Conclusion}
We resume the main results of our paper. We presented review of the map construction which provides the correspondence rule between the operators acting in a Hilbert space and the functions of some variables. The construction of the map is based on using the pair of operators depending on these variables called quantizer and dequantizer operators. These operators give possibility to consider the operators acting in Hilbert space as a vectors and the functions (symbols of operators) as components of these vectors which are coordinates in specific basis in the linear space. Then the pair of quantizer and dequantizer operators correspond to the bases in the linear space providing the possibility to consider any operator in Hilbert space as a vector in the linear space which is given if the components of the vector in this basis are known. Several examples of such construction were considered. Wigner-Weyl symbols of operators, symplectic tomographic symbols, photon number tomographic symbols were studied using the pairs of corresponding quantizer-dequantizer operators. In this formalism the evolution equation for quantum states can be written in the form of kinetic equation for the probability distribution \cite{AmosovKorennoy}. The evolution of tomograms for different quantum systems, both finite and infinite dimensions was considered in \cite{Kishore}. A method based on tomography representation to simulate the quantum dynamics was applyed to the wave packet tunneling of one and two interacting particle in \cite{Arhipov}. Non-orthogonal bases of projectors on coherent states were introduced to expand Hermitean operators acting on the Hilbert space of a spin $s$ in \cite{Amiet}.

In our work we discuss the kernels of star-product of symbols in the case of mentioned examples. These kernels generalized the known Gr\"oenweld kernel which provides the associative  product of the Wigner--Weyl symbols \cite{Groen}. So, we consider in this article the examples of infinite dimensional Hilbert spaces. The analogous construction exists for finite dimensional Hilbert spaces corresponding to the states of spins and qudits \cite{DodPLA}, \cite{OlgaJetp}, \cite{Bregence}, \cite{Wiegert2}, \cite{Castanos}, \cite{Oktavio}, \cite{Fillipov}, \cite{Fedorov}. In \cite{Elizabett} a class of deformed products for spin observables is presented. A review of different quantum-optical states defined in finite-dimensional Hilbert space of operators which have a discrete spectrum was considered in \cite{Miranovich}. The methods of star-product, tomography and probability representation of quantum mechanics applied to different problems of quantum phenomens in  \cite{Mosna}-\cite{Xiao}. The quantization method based on star-product of functions can be applied for studying the process in different areas of physics.

\end{document}